\documentclass[onecolumn]{IEEEtran}
\usepackage{cite}
\usepackage{amsmath,amssymb,amsfonts}
\usepackage{algorithmic}
\usepackage{textcomp}
\usepackage[utf8]{inputenc}
\usepackage[T1]{fontenc} 
\usepackage{hyperref} 
\usepackage{mathtools}
\usepackage{graphicx}
\usepackage{tabularx}
\usepackage{subfigure}
\usepackage[table]{xcolor}
\graphicspath{{images/}}
\usepackage{dblfloatfix}
\usepackage{fixltx2e}
\usepackage{multirow}
\usepackage[justification=centering]{caption}
\usepackage{array}	
\usepackage{algorithm}

\begin{document}

\title{Fog Architectures and Sensor Location Certification in Distributed Event-Based Systems~$^{\dagger}$}

\author{\IEEEauthorblockN{F\'atima Castro-Jul}
\IEEEauthorblockA{AtlantTTic Research Center, University of Vigo, Spain.
Email: avilas@det.uvigo.es}\\
\and
\IEEEauthorblockN{Rebeca Díaz Redondo}
\IEEEauthorblockA{AtlantTTic Research Center, University of Vigo, Spain.
Email: rebeca@det.uvigo.es}\\
\and
\IEEEauthorblockN{Ana Fern\'andez-Vilas}
\IEEEauthorblockA{AtlantTTic Research Center, University of Vigo, Spain.
Email: avilas@det.uvigo.es}
and
\IEEEauthorblockN{ Sophie Chabridon}
\IEEEauthorblockA{AMOVAR, T\'el\'ecom SudParis, CNRS, Universit\'e Paris-Saclay, 91000 \'Evry, Francn}
and
\IEEEauthorblockN{Denis Conan}
\IEEEauthorblockA{AMOVAR, T\'el\'ecom SudParis, CNRS, Universit\'e Paris-Saclay, 91000 \'Evry, France}}
\maketitle

\begin{abstract}
Since smart cities aim at becoming self-monitoring and self-response systems, their~deployment relies on close resource monitoring through large-scale urban sensing. The~subsequent gathering of massive amounts of data makes essential the development of event-filtering mechanisms that enable the selection of what is relevant and trustworthy. Due to the rise of mobile event producers, location information has become a valuable filtering criterion, as it not only offers extra information on the described event, but also enhances trust in the producer. Implementing mechanisms that validate the quality of location information becomes then imperative. The lack of such strategies in cloud architectures compels the adoption of new communication schemes for Internet of Things (IoT)-based urban services. To~serve the demand for location verification in urban event-based systems (DEBS), we have designed three different fog architectures that combine proximity and cloud communication. We have used network simulations with realistic urban traces to prove that the three of them can correctly identify between 73\% and 100\% of false location claims.

\end{abstract}


\section{Introduction}

Smart cities are intelligent and interconnected environments where resources are constantly monitored in order to, among other objectives, assess their performance and ensure an appropriate reaction in case of an incident. From transportation to safety or energy management, the list of services that can benefit from obtaining real-time information to enhance citizens' experience is almost endless~\cite{mone2015new}. Monitoring a whole city implies the extensive sensing and gathering of large amounts of data. The emergence of the Internet of Things (IoT) eases this task by enabling the interconnection of sensors embedded in any device, vehicle, or object.

Since they provide time and space decoupling, distributed event-based systems (DEBS)~\cite{Eugster2003} enable flexible communication in dynamic and heterogeneous urban scenarios. As a result, they have become a useful interaction mechanism for IoT-based smart-city services~\cite{antonic2014}. Moreover, they support event filtering, data aggregation, and scalability, which are essential in the development of large-scale IoT-based services~\cite{cugola2012processing}. Event filtering is of paramount importance in large-scale urban sensing, where~an enormous quantity of data are generated. By choosing their event subscriptions, consumers can select the most relevant and trustworthy data available to improve their environment awareness in a reliable way. Therefore, event filtering is not limited to the event notification content, but also extends to its quality, evaluated using additional knowledge provided by event producers and event processing agents~\cite{bellavista2012survey}. 

When considering filtering criteria, location is one of the first that come to mind. The inclusion of geographical information on event notifications allows subscriptions based on areas of interest. Thus, it helps consumers to select notifications that are relevant to them. Additionally, it reinforces trust in producers. Even though location information may not be significant in every notification, the fact that producers provide extra information about their whereabouts increases their reliability. As a result, implementing location evaluation mechanisms improves trust both on the event notification source and on the notification itself. 

Location assessment mechanisms differ according to the producers' mobility and the DEBS architecture. Static event producers are known to be in a certain location and, therefore, it is easy to detect fake location claims coming from them. However, when it comes to mobile producers, location assessment becomes more difficult. The importance of mobile producers has considerably grown over the last ten years due to the emergence of participatory sensing~\cite{burke2006} and mobile crowd-sensing paradigms~\cite{guo2014}. Producers that move through the city and gather information about different areas are a promising alternative to building a costly fixed sensing and communication infrastructure to cover a whole city. As a result, mobile clients and subsequent location issues need to be considered when designing urban DEBS. In classical cloud architectures, an overlay of brokers in the Cloud lets producers and consumers communicate, placing blind confidence in the locations they claim. A~possible solution is to switch from this architecture to a proximity-based totally ad hoc network that ensures producers' direct communication with consumers in their very area~\cite{castro2018collaboratively}. This is a good solution for small and densely populated urban areas, but it becomes expensive when producers and their subscribed consumers are far away from each other. Another alternative is to migrate part of the data processing from the Cloud to the edges of the network. Thus, location is verified on the edges, in~the devices' proximity, while a cloud architecture for widespread area communication is maintained. Extending cloud architectures closer to the users is not a novel idea, it is the basis of the Fog Computing paradigm~\cite{bonomi2012,perera2017fog}. 

The capabilities of fog architectures for sensor location certification in urban DEBS remain largely unexplored. In order to 
evaluate their potential, we have designed and assessed through simulation three different architectures based on fog computing that combine cloud and proximity communication. The architectures are aimed at improving the quality of urban notifications and enable wide-area communication while benefiting from proximity-based location verification on network edges. To do so, we enrich traditional cloud architectures with short-range communication schemes that enable peer-based location verification. This proposal evolves from our works on quality of context information in DEBS~\cite{Lim2016} and on distributed proximity-based collaboration with peer devices~\cite{castro2018collaboratively}. The~architectures, albeit briefly presented in our previous work~\cite{castro2017combining}, are extensively studied in this paper. Moreover, their performance is assessed using network simulations with realistic urban traces. 

This paper is organized as follows. First, Section~\ref{sec:background} presents an overview of how location verification is dealt with in urban-sensing architectures. Then, Section~\ref{sec:scenario} analyzes the scenario at which our approach is targeted, and Section~\ref{sec:system} outlines the basics of our DEBS system. Our proposed architectures are described in Section~\ref{sec:architecture}, while their simulation and evaluation are discussed in Sections~\ref{sec:simulation} and \ref{sec:evaluation}. Finally, a conclusion and some guidelines for future work can be found in Section~\ref{sec:conclusion}.

\section{Background} \label{sec:background}
 The high potential of mobile sensor-enabled devices as a powerful tool to monitor their surroundings has led to the emergence of participatory~\cite{burke2006} and crowdsensing systems~\cite{ganti2011, guo2014}, which~leverage user-generated information to form collaborative knowledge about urban areas. These~systems enhance the quality and credibility of urban sensing and, therefore, have become a major player in IoT-distributed event-based platforms~\cite{antonic2016mobile}. However, the prevalence of mobile event producers poses new challenges in urban-sensing schemes.
 
On the one hand, the possibility of exploiting an ever-growing number of mobile devices results in a massive increase in the quantity of collected data. As a result, developing filtering mechanisms becomes imperative. The focus of those may be reducing the transmission of unnecessary information on the publishing side~\cite{antonic2014,antonic2016mobile,castro2018collaboratively} or supporting selective subscriptions on the consuming side, based on criteria such as trust on the producer and information quality~\cite{Lim2016}.

On the other hand, sensing platforms need to handle inherent device mobility. This includes dealing with unequal area coverage due to human mobility patterns~\cite{antonic2014}, translating information on producers in different locations~\cite{dang2017data}, and analyzing location claims that are continuously changing. Since sensed information is often geotagged for contextualization, verifying location claims is of paramount importance to ensure adequate subscription management and to reinforce trust in the sensing system. Thus, the location verification problem relates to both of the mentioned challenges in participatory sensing: it arises from the high mobility of producers and, considering that location is a major filtering criterion, it significantly affects effective data dissemination. The task of designing participatory DEBS with location verification support varies depending on the different system architectures, whose main characteristics are described next.

DEBS are typically based on three components~\cite{Eugster2003}: producers ($P$), consumers ($C$), and brokers~($B$). Producers advertise their contribution to the system and then publish events that match their advertisement. Consumers subscribe to events of their interest and receive them. Brokers handle consumer subscriptions in order to route event notifications appropriately. Every client (consumer or producer) is connected,  at most, to one broker at the same time. We distinguish between three different system architectures: cloud (Section~\ref{subsec:background-cloud}), cloudless (Section~\ref{subsec:background-cloudless}), and fog (Section~\ref{subsec:background-fog}).

\subsection{Cloud Architectures}\label{subsec:background-cloud}
Traditional cloud architectures~\cite{antonic2014,antonic2016mobile} consist of an overlay of brokers in the Cloud that communicates producers and consumers located anywhere. Figure~\ref{fig:architectures-related}a depicts a cloud-based participatory sensing with producers ($P_i$) and consumers ($C_j$) in different city sections. There is also an overlay of interconnected brokers ($B_k$) that allows communication over the whole area and whose internal organization is transparent to the clients. The clients (producers and consumers) first connect  to a central server, which assigns them to an access broker. The access broker can be anywhere, and may serve other clients from different locations. It does not know the clients' position and has no means to verify whether they are claiming false locations. As a result, the access broker places blind trust in the locations submitted by producers.\\

\subsection{Cloudless Architectures}\label{subsec:background-cloudless}
\begin{figure}[H]
\centering
\begin{minipage}[b]{0.8\textwidth}
\centering
\includegraphics[width=1\textwidth]{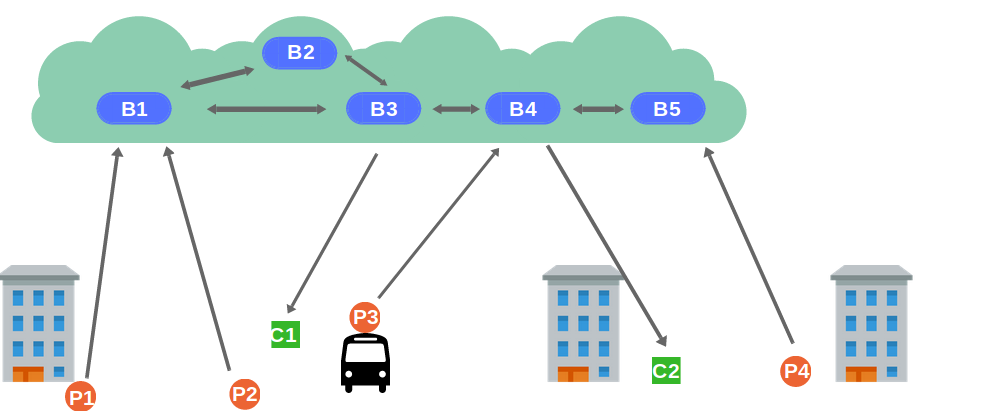}
\caption{Cloud architecture\label{subfig:cloud-architecture}}
\end{minipage}%
\\
\begin{minipage}[b]{0.8\textwidth}
\centering
\includegraphics[width=1\textwidth]{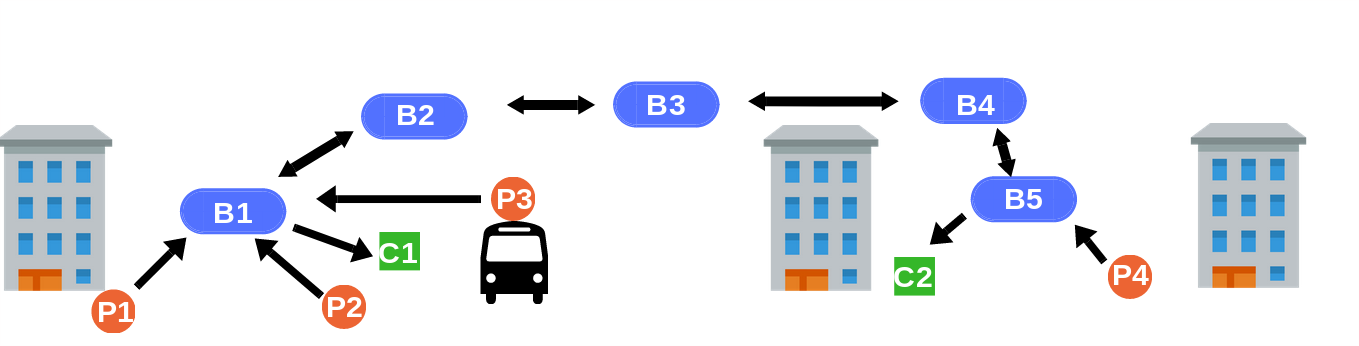}
\caption{Cloudless infrastructure architecture\label{subfig:cloudless-infrastructure-architecture}}
\end{minipage}%
\\
\begin{minipage}[b]{0.8\textwidth}
\centering
\includegraphics[width=1\textwidth]{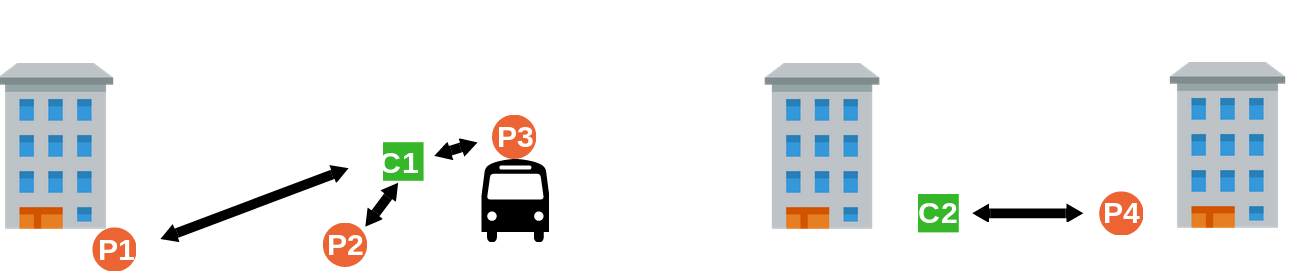}
\caption{Ad hoc architecture\label{subfig:ad-hoc-architecture}}
\end{minipage}%
\caption{Participatory sensing architectures.\label{fig:architectures-related}}
\end{figure}

Moving the whole sensing system to the clients' proximity ensures producers communicate only with those in their actual whereabouts. By dispensing with the Cloud and relying solely on short-range communication technologies, false location claims become easily detectable. Cloudless architectures may consist of a broker infrastructure placed all over the urban area that needs to be covered (Figure~\ref{fig:architectures-related}b). Access brokers ($B_k$) are always near their clients and, therefore, can verify whether their location claims are accurate. Brokerless or ad hoc architectures are another alternative (Figure~\ref{fig:architectures-related}c). Clients ($P_i$ and $C_j$) communicate directly with each other and form opportunistic networks to deliver event notifications to those interested~\cite{castro2018collaboratively}. In this scenario, every client takes part in the verification of notifications created in their proximity. Both the architectures work well in small and, in the ad hoc case, densely populated zones. However, they are costly to deploy in wide areas and do not enable communication with clients outside of it.

\subsection{Fog Architectures}\label{subsec:background-fog}
The Fog Computing paradigm advocates the establishment of an intermediate layer between end devices and traditional cloud servers that provides computation, storage, and networking services~\mbox{\cite{bonomi2012,perera2017fog}}. It has emerged as a useful tool for IoT architectures since it provides mobility and heterogeneity support~\cite{yannuzzi2014}, low latency, and scalability~\cite{Hong2013}. Moving part of the data processing closer to the ground increases the system awareness about clients' locations. Moreover, since part of the architecture is still in the Cloud, widespread area coverage is guaranteed. There exist different alternatives in the design of fog-computing architectures, where the intermediate layer can be placed either remotely or physically. The recent emergence of fog architectures requires the development of specific trust and security strategies targeted at their distributed structure~\cite{zhang2018security}.

\subsection{Our Contribution}
To the best of our knowledge, there exist no distributed solutions for location verification support in event-based sensing systems. The lack of such mechanisms prevents the implementation of reliable location-based subscriptions and reduces the confidence that consumers place in event producers and, thus, in the system itself. As a consequence, the development of urban services based on information gathering through participatory event-based schemes is compromised. Hence, it is imperative to analyze the potential of fog architectures for location verification in DEBS.

Our contribution focuses on employing proximity-based communication to verify producers' location. Proximity-based communication connects devices that are placed in the same area, at a distance from each other that allows them to be reached using wireless short-range communication technology, such as Bluetooth or ad hoc WiFi. This kind of connection diminishes the possibility of producers to fake their location beyond just a few meters. If a producer that claims to be in a particular area cannot be reached by other devices known to be there, we can conclude that it is not providing its actual location. Following this principle, we have designed three different DEBS architectures that support location verification: fixed-brokers architecture, assigned-brokers architecture, and collaborative architecture. Before detailing our location verification proposal and introducing the architectures,  in the next section we present the scenario we used for illustrative~purposes.

\section{Scenario}\label{sec:scenario}
We have targeted our scenario at a participatory sensing DEBS with high producer mobility and periodic sensing-data publications. We chose an environmental variable that can be detected by merely using a smartphone, and presents multiple applications: noise monitoring. Direct measurements are employed for noise-pollution assessment, which is a topic of concern due to its consequences on citizens' health and wellbeing. Moreover, noise measurements can lead to the inference of different events, such as traffic jams, roadworks, or crowds. A pub/sub system where users can receive geolocated noise data helps them to avoid highly polluted or congested areas of the city by choosing alternative routes. This kind of system also serves as a detection mechanism for events such as concerts or demonstrations. In order to provide consumers with precise event positioning, it is essential to provide reliable location data. 

To avoid the high cost of citywide infrastructure deployment, noise measurements can be performed using mobile sensors carried by vehicles or users. The latter is especially interesting since smartphones' microphones have been proven to successfully act as noise sensors~\cite{d2013}. Therefore, no~dedicated infrastructure is necessary. Having mobile devices engaged in dynamic and interactive networks to share local information makes this a participatory sensing scenario~\cite{burke2006}.

Noise measurements are automatically issued every few minutes. Thus, possible events can be quickly detected, and subscribers frequently receive fresh data. Using a fog architecture allows users to receive noise-measurement event notifications anywhere. Therefore, they can, for instance, be aware of incidents in their place of residence even when they are away, or discover if an area in a different city they want to move or travel to is noisy.

The same system can be adapted to other urban variables, such as air pollution. Furthermore, it~can be targeted at systems where producers are not mere information receivers but working sensors that can prioritize their activity \cite{costa2018twittersensing} or change their configuration parameters \cite{costa2017fuzzy} when required by the circumstances.

\section{DEBS with Location Verification Support} \label{sec:system}
Our goal is to improve event notification quality by increasing trust in the event's source location. In order to focus our attention on producers' mobility, we assume the broker and consumer structure to be fixed. Producers advertize locally and consumers subscribe globally. Thus, the broker network is always aware of subscription needs and puts consumers in communication with the appropriate producers. Simple routing is employed: each subscription leads to the installation of a direct spanning tree that routes notifications toward the consumer.

Location information is inserted as an attribute in every event notification. If it can be verified, notifications also include location certification. The location attribute can be verified either on the producer or on the broker, but it is always the broker, assumed to be trustworthy, which decides whether to provide the certification. Consumers may choose to subscribe only to events with a certified location.
The verification process consists in assessing the location attribute using either extra information provided by the producer to support the location claim or the collaboration of other peers in the area. The certification process consists of adding a location certificate in the event notification to state the veracity of the location claim. An event notification is certified if the location claim has been verified. The verification and certification processes ensure that at least one agent has reviewed the location attribute provided by the producer and assessed it. As a result, trust in the notification source is increased.

However, we are only interested in producer trust in order to extend it to the notification content. Unlike other trust-based systems~\cite{josang2007survey}, ours is not based on reputation, as we neither implement a rating mechanism nor maintain a history of contributions. The trust our system provides is actually related to the notion of provision trust~\cite{josang2007survey} and refers to the system's capability to differentiate reliable information to serve consumers' subscriptions.

In this section, we describe the distinct characteristics of a DEBS with location verification support: the role of producers and brokers in the verification, the definition of the location attribute, and the management of mobility. Moreover, we indicate how some of the features differ according to the three architectures we propose in this paper, which are described in detail in Section~\ref{sec:architecture}: the fixed-brokers architecture, the assigned-brokers architecture, and the collaborative architecture. 

\subsection{Producers' Role}
For the sake of simplicity and without impairing generality, notifications contain attributes as sets of triples: ($n_{1,i}$, $n_{2,i}$, $v_{i}$), where $n_{1,i}$ and $n_{2,i}$ are strings that uniquely identify the attribute, and $v_i$ is the value of the attribute. Since an attribute can be associated not only with its value but also with metadata, it may be described by multiple triples, e.g., $(\textsf{location}, \textsf{value}, \textsf{X})$, $(\textsf{location}, \textsf{certified}, \textsf{Y})$. For instance, the location attribute is inserted in the notification as the triple $(\textsf{location}, \textsf{value}, \textsf{X})$, where \textsf{X} is a variable that represents the value of the location. Producers include this latter triple in every notification. Moreover, the publishing process may also involve the verification of the location attribute and the incorporation of extra information to assist the broker in the verification process. In~Algorithm~\ref{alg:publish}, function PrepareExtraLocationInformation deals with the tasks of verifying location and arranging extra location data in case they are considered. It returns the arranged information, which~function PrepareNotification inserts in the notification together with the location triple. Both~of these functions vary their operation according to the architectures. Their different definitions are presented in Section~\ref{sec:architecture}.

Since a producer can disconnect and reconnect, its architecture is supplemented with a queue of notifications where they are stored when no access broker is reachable. Queued notifications are transmitted to the access broker that the client connects next.

\begin{algorithm}[H]
    \caption{: at producer $P$, publish \label{alg:publish}}
    \begin{algorithmic}[1] 
   \STATE Publish(Notification $n$)
         \STATE $lc \leftarrow$ PrepareExtraLocationInformation()  
      \STATE $n \leftarrow$ PrepareNotification($n$,$lc$) \COMMENT{Adds location value and extra information, if any}
       \IF {connected to an access broker}
          \STATE send $n$ to the access broker
      \ELSE
          \STATE store $n$ in the local queue
        \ENDIF
    \end{algorithmic}
\end{algorithm}

\subsection{Location Attribute}
The location attribute may consist of an explicit representation using coordinates or an identifier of a specific area, neighborhood, or city. The former supports more precise location data but implies more expensive event processing. Considering that the areas are delimited by short-range communication technology, we considered that area identifiers were precise enough in this work. We~assumed the existence of a location ontology that is known by all the agents in the system. Moreover, producers incorporate a geolocation mechanism that allows them to identify their own position and describe it with a valid location value.

The location ontology can include a hierarchical structure that allows the definition of different levels of location granularity and notification visibility. 

\subsection{Brokers' Role}
Brokers act not only as notification routers but also involve event processing agents. A common one for the three proposed architectures is a translate event processing agent~\cite{etzion2011event}, in charge of enriching the verified notifications with a location certificate in the form of the attribute triple: $(\textsf{location}, \textsf{certified}, \textsf{Y})$, where the first two elements describe the attribute and \textsf{Y} can take two possible values: \{$true, false$\}.  To establish the veracity of the location claim, the broker may rely on the extra information provided by the producer in the notification. The nature and the treatment of this information vary in the different architectures. In the first two architectures (fixed-brokers architecture, assigned-brokers architecture), there exists a filtering agent that identifies the notifications whose location is the same as the broker's. Additionally, brokers in the assigned-brokers architecture and the collaborative architecture integrate an agent whose role is to check the information provided by the producers to support their location claim and to remove it from the notification. The function VerifyLocation is in charge of processing the extra information, removing it from the notification, and returning a value that indicates whether the location claim has been verified or not. The definition of the function differs according to the different architectures, as detailed in Section~\ref{sec:architecture}.

Algorithm~\ref{alg:handle-notification} depicts the notification handling process.

\begin{algorithm}[H]
    \caption{\small{: at broker $B$, handle local notifications \label{alg:handle-notification}}}
    \begin{algorithmic}[1] 
      \STATE HandleLocalNotification (Producer $P$, Notification $n$)
         \IF {$B$ is the access broker of $P$} 
           \STATE $locOk \leftarrow$ VerifyLocation($n$)
           \IF{$locOk$}
             \STATE insert triple $(\textsf{location},\textsf{certified},true)$
           \ELSE 
             \STATE insert triple $(\textsf{location},\textsf{certified},false)$
         \ENDIF
         \ENDIF
         \STATE Forward $n$ to interested neighbor brokers and local consumers
    \end{algorithmic}
\end{algorithm}

\subsection{Mobility}
When brokers are assigned to geographical areas, a change in physical location also implies a change in broker connection. Regardless of their location, producers publish periodic notifications that are processed by the access broker to which they are connected at that time. Thus, mobility handling is transparent for the producers. This concept of mobility corresponds to the physical mobility defined by Fiege et al.~\cite{fiege2003supporting}. In the first two proposed architectures, mobility is handled by the local broker, installed on every client. It is in charge of establishing connections with border brokers and delivering them the notifications. It becomes aware of a location change when a broker connection is no longer~available. 

In the fixed-brokers architecture, this happens when the producer moves out of reach of the broker communication sphere. Then, the producer listens for beacons from another border broker to which to connect. In the assigned-brokers architecture, the connection is canceled when the producer is no longer able to send valid periodic notifications. At that point, the producer becomes aware of the location change and sends a new connection request with the producer's location coordinates. 

Event notifications published by disconnected producers are queued locally. When a new connection is established, they are issued to the new border broker. If the location attribute of the queued notifications is different from the new broker's location, the notifications are distributed but are not certified because a broker cannot certify a different location from its own.

The situation is different in the collaborative architecture. There, changes in broker connection do not imply location changes and vice versa. Since brokers are not assigned to a certain area, there are no limitations on the notifications they can consider for certification. As long as the location meets the criteria to be considered collaboratively decided (defined in Section~\ref{subsec:architecture-collaborative}), it can always be certified regardless of changes in the broker connection or the producer location.

\subsection{Threat Model}\label{subsec:threat-model}
We consider the case where a deceitful producer decides to include a false location claim in a publication. This attack has been referred to as location spoofing~\cite{saracino2018practical}. We assumed that there may be more than one misbehaving producers but that they act independently and never collude or conspire.

We also assumed that all  brokers are trustworthy and that they accurately certificate the publications. Moreover, we did not consider the involvement of third parties and we did not deal with any other type of attack.

\section{Architectures}\label{sec:architecture}
In this section, we present three different fog architectures for a DEBS system with location verification support and we outline their main features.

\subsection{Fixed-Brokers Architecture}

\begin{figure}[H]
\centering
\includegraphics[width=0.8\textwidth]{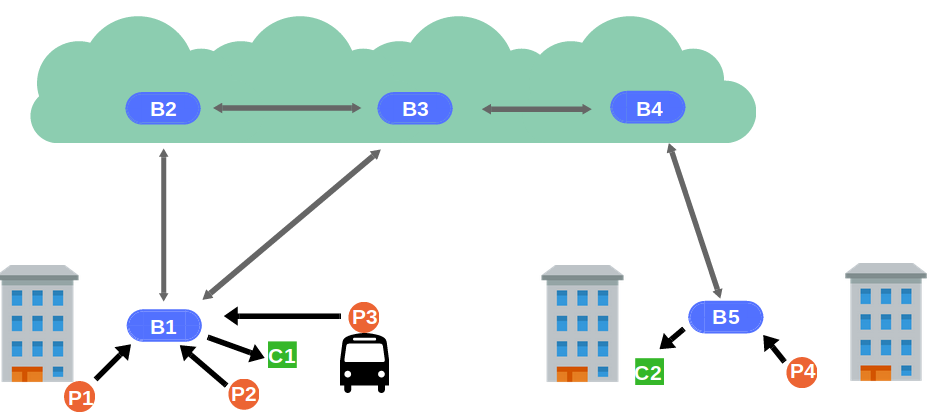}
\caption{Fixed-brokers architecture.\label{fig:fixed-brokers-architecture}}
\end{figure}

The first architecture is shown in Figure~\ref{fig:fixed-brokers-architecture}. There exist two levels of brokers: border brokers (B1, B5) and inner brokers (B2, B3, B4). While inner brokers are in the Cloud, border brokers are placed around the city covering different areas. They are fixed and, therefore, they are assigned a fixed location value. Access brokers are connected to inner brokers to enable event distribution. Clients cannot directly interact with inner brokers, whose overlay structure is transparent to them. Clients in the same area are connected to the same access broker using a short-range transmission technology that ensures the veracity of the clients' location. Thereby, in this architecture location verification is straightforward. There is no verification on the producer (Algorithm~\ref{alg:producer-fixed-brokers}), and the access broker certifies the location of every event notification whose location attribute coincides with its own location (Algorithm~\ref{alg:broker-fixed-brokers}).

\begin{algorithm}[H]
    \caption{\small{: at producer $P$, prepare extra location information (Fixed-brokers architecture) \label{alg:producer-fixed-brokers}}}
    \begin{algorithmic}[1] 
    \STATE PrepareExtraLocationInformation()
       \RETURN $\emptyset$
    \end{algorithmic}
\end{algorithm}

In this architecture, connections are simple. Brokers advertize themselves through beaconing using short-range communication technology. When a producer that is not connected to any broker is reached by a beacon, it establishes a connection with the sending broker. The producer and the broker exchange periodic messages to maintain the connection. When they stop receiving these messages, they consider the connection as broken. The broker deletes the producer's advertisements and the producer starts paying attention to beacons to establish a connection with another broker.

\begin{algorithm}[H]
    \caption{\small{: at broker $B$, verify location (Fixed-brokers architecture) \label{alg:broker-fixed-brokers}}}
    \begin{algorithmic}[1] 
    \STATE VerifyLocation(Notification $n$)
     \STATE $location  \leftarrow \{Y | \{\textsf{location},value,Y\} \in n\}$
     \IF {$location == B's$ $location$}
       \RETURN true
      \ELSE
          \RETURN false
     \ENDIF
    \end{algorithmic}
\end{algorithm}

\subsection{Assigned-Brokers Architecture}

\begin{figure}[H]
\centering
\includegraphics[width=0.8\textwidth]{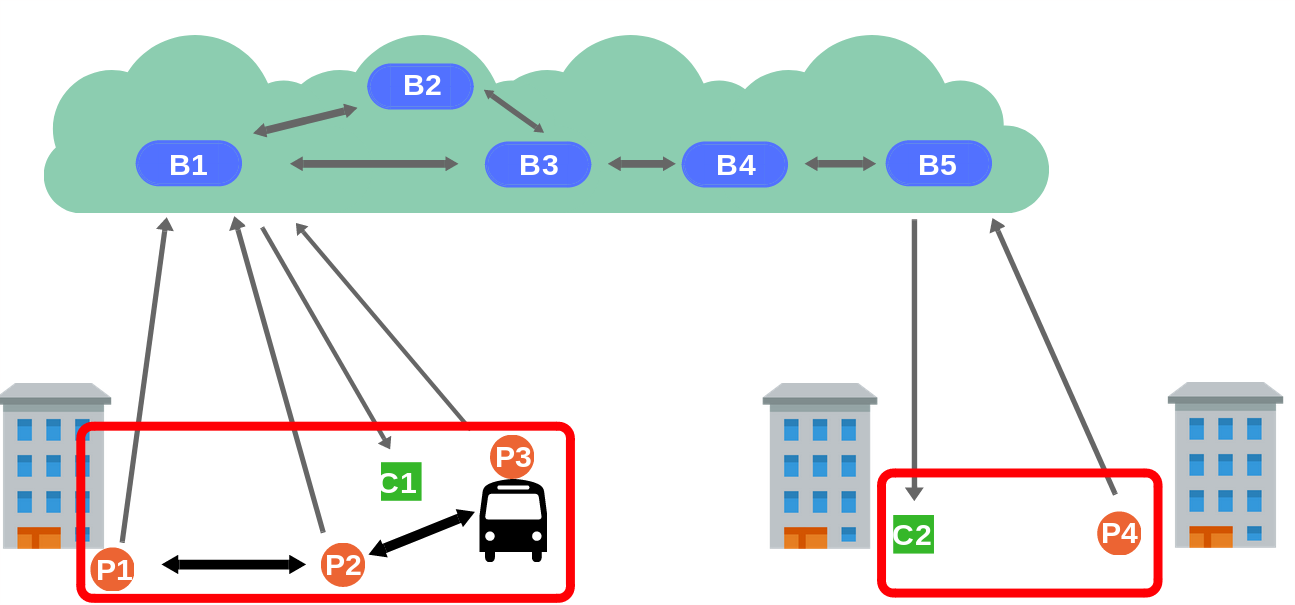}
\caption{Assigned-brokers architecture.\label{fig:assigned-brokers-architecture}}
\end{figure}

Having a hybrid architecture that combines fixed and mobile agents diminishes the infrastructure requirements of deploying a purely fixed sensor architecture. However, the cost of setting up fixed brokers may still be high. As a result, we have designed a second architecture that also exploits location verification through proximity-based communication, but gives up the requirements for fixed infrastructure. Figure~\ref{fig:assigned-brokers-architecture} displays the same DEBS network shown in Figure~\ref{fig:fixed-brokers-architecture} but with all the brokers in the Cloud. By considering location-based network partitions, we can organize client connections to brokers to maximize the probability of having all the clients in an area connected to the same broker. Then, brokers can still be in charge of producers' location verification even though they cannot communicate with them using short-range transmission technology. Each broker is assigned an area and maintains a list of registered producers, updated with every publication received. Producers obtain the IP address of their access brokers by providing their location to a discovery service. To~maintain the connection, producers are required to periodically send valid notifications to the broker.

When a producer sends a notification, it includes the list of other producers in the area with whom it has established a short-range communication link and who are connected to the same broker (Algorithm~\ref{alg:producer-assigned-brokers}). Producers claiming a false location are not able to provide a valid neighbor list. Then, the broker compares the neighbor list provided in the received notification with the list of devices in the area it maintains. The publication is certified if it satisfies the requirements detailed below. If the publication is certified, the sender is included in the list of registered producers in that area.

\begin{algorithm}[H]
    \caption{\small{: at producer $P$, prepare extra location information (Assigned-brokers architecture) \label{alg:producer-assigned-brokers}}}
    \begin{algorithmic}[1] 
    \STATE PrepareExtraLocationInformation()
        \RETURN neighborList  
    \end{algorithmic}
\end{algorithm}

Relying solely on mobile producers involves extra challenges. Producers on area edges are likely to be reached by beacons from producers in a different area. Therefore, information on neighbors from nearby regions need to be considered when assessing the location of every publication. As a result, every broker should exchange its list of registered producers with the brokers assigned to contiguous~areas.

\subsubsection*{Verification Strategies}\label{subsubsec:verification}
Two different verification strategies are proposed: the complete-list strategy (CLS) and the nonempty list strategy (NLS). In the first one, the neighbor list provided by the producer needs to satisfy two requirements. 
First, it has to include every producer registered with the broker assigned to that area. This condition requires areas small enough for 
producers to find every one of their peers registered to the same broker with high probability.\\
Second, the list cannot include extra neighbors that are neither registered with the broker nor with the brokers assigned to adjacent areas. Otherwise, producers may include a neighbor list with as many producer identifiers as possible in the hope that some of them are actually registered with the broker. These requirements can be written as:
\begin{enumerate}[leftmargin=*,labelsep=4.9mm]
\item $\forall b \in P_{B},$ $b \in$ \textit{NP}
\item $\forall b \in$\textit{NP}$,$  $b \in P_{B} \cup P_{NB}$
\end{enumerate}
where \textit{NP} is the list of neighbors provided by the producer, $P_{B}$ is the list of producers registered with the broker assigned to that area, and $P_{NB}$ is the list of producers registered with brokers assigned to contiguous areas.

CLS is not effective in situations where the producer has not found any neighbors and there are no producers registered with the broker either, i.e., both $NP$ and $P_{B}$ are empty. In this case, the broker certifies the notification regardless of the veracity of the location claim.

To resolve this issue, we propose a second verification strategy (NLS), where the producer is required to provide a neighbor that is registered with the broker, i.e., $b \in NP \cup P_{B}$. However, this~strategy is not optimal either. It could lead to the certification of notifications with false location claims that include a long list of producers, whereof one happens to be registered with the broker. This is problematic in scenarios where a producer can maintain a list of every producer it has ever known and use it as its current neighbor list.

 As a result, CLS is better suited to scenarios with high device density where there are producers in every area and, therefore, assigned to every broker. NLS is targeted at scenarios where the density of devices is low and where no producer is assumed to have global knowledge about peers in the whole system. We include both the strategies in our simulation analysis.
 
Algorithms~\ref{alg:broker-assigned-brokers-1} and \ref{alg:broker-assigned-brokers-2} describe the location verification procedure in the broker for CLS and NLS,~respectively. 

\begin{algorithm}[H]
    \caption{\small{: at broker $B$, verify location (Assigned-brokers architecture, CLS) \label{alg:broker-assigned-brokers-1}}}
    \begin{algorithmic}[1] 
    \STATE VerifyLocation(Notification $n$)
       \STATE $NP \leftarrow$ $neighborList \in n$
       \STATE remove $neighborList$ from $n$
       \STATE $location  \leftarrow \{Y | \{\textsf{location},value,Y\} \in n\}$
      \IF {$location == $B's location $\AND (\forall b \in P_{B},$ $ b \in NP) \AND (\forall b \in NP,$ $ b \in P_{B} \cup P_{NB}$)}
          \RETURN true
       \ELSE  \RETURN false
       \ENDIF
    \end{algorithmic}
\end{algorithm} 

\begin{algorithm}[H]
    \caption{\small{: at broker $B$, verify location (Assigned-brokers architecture, NLS) \label{alg:broker-assigned-brokers-2}}}
    \begin{algorithmic}[1] 
    \STATE {VerifyLocation}{Notification $n$} 
     \STATE $NP \leftarrow$ $neighborList \in n$
       \STATE remove $neighborList$ from $n$
       \STATE $location  \leftarrow \{Y | \{\textsf{location},value,Y\} \in n\}$
      \IF {$location == $B's location $\AND (\exists b \in NP,$ $b \in P_{B})$}
           \RETURN true
       \ELSE  \RETURN false
       \ENDIF 
    \end{algorithmic}
\end{algorithm}

\subsection{Collaborative Architecture}\label{subsec:architecture-collaborative}

\begin{figure}[H]
\centering
\includegraphics[width=0.8\textwidth]{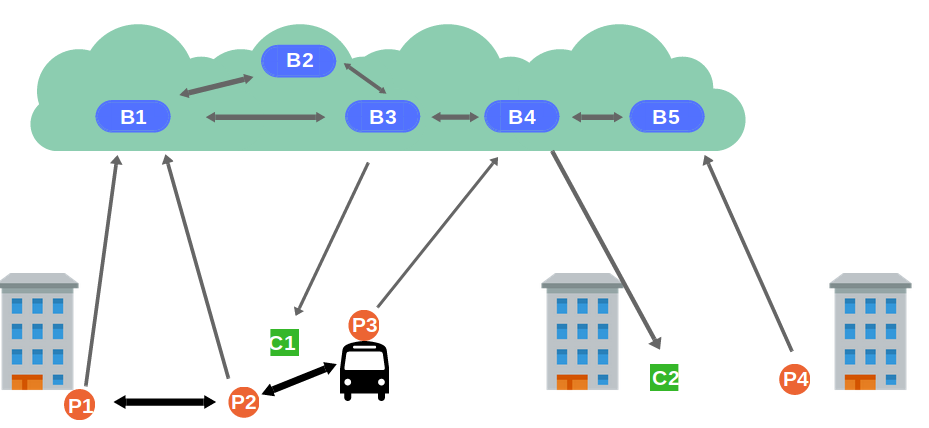}
\caption{Collaborative architecture.\label{fig:collaborative-architecture}}
\end{figure}

To reduce the burden of having brokers responsible for managing producers' location, we~have designed a third architecture that minimizes the broker role by placing location verification on the producer side. This architecture relies on traditional cloud architecture. Producers can be connected to any border broker, which may be different from the one their close neighbors are connected to. As depicted in Figure~\ref{fig:collaborative-architecture}, this architecture incorporates short-range connections between nearby producers so they can communicate with each other and collaboratively decide the location they tag their notifications with. Every producer is equipped with a local broker that is in charge of handling the notification and the location provided by the producer. Then, it communicates with local brokers of nearby producers. Finally, it sends the notification to a border broker in the cloud with the collaboratively decided location. In our previous work~\cite{castro2017combining}, we proposed the use of a consensus strategy as the mechanism for collaborative location decision. However, due to the high mobility of the producers, which arrive at or leave an area at any time, the possibility of forming a stable group of nodes that satisfies the requirements to take part in a consensus agreement is low. As a result, we have decided to employ an alternative collaborative strategy: neighbor polling.\\

 Every time a publication is created, the local broker polls nearby producers by sending a short-range broadcast message. Thus, collaboration is restricted to nodes in the same physical area. The local broker waits for a certain time, computes the location values received and includes in the publication the decided value. The location value proposed locally is also included in the poll. 
The~decided location is the most repeated location in the poll, as long as at least one neighbor answer was received and there is no tie between the most voted locations. This requirement can be written as:
\begin{enumerate} [leftmargin=*,labelsep=4.9mm]
\item $RL > 1$
\item $\exists! b \in R,$ $ b = max(R)$
\end{enumerate} 
where $RL$ is the list of received locations plus the one provided by the local broker, which consists of entries formed by two triples: 
$e = \{( \textsf{location}, \textsf{value},X), (\textsf{numberOfReplies}, \textsf{value}, Z)\}$.

$R$ is the list that includes the number of poll replies for every location. That is:\\ $R \leftarrow \{Y|\{( \textsf{location}, \textsf{value},X), (\textsf{numberOfReplies},*, Y)\} \in RL \}$. 
  
Publications issued to the broker include an indication of whether the location has been collaboratively decided, in the form of the triple: $(\textsf{location},\textsf{collaborativelyDecided}, \textsf{X})$, where \textsf{X} is a boolean variable whose value is $true$ if the collaboration has been successful. In case no poll replies are received, or there is a tie between the most voted locations in the poll, the result is undecided. Then, the location proposed locally is included in the notification together with an indication of unverified location, i.e., a triple where \textsf{X} takes the value $false$. Algorithm~\ref{alg:producer-collaborative-architecture} describes the location verification procedure for this strategy.
\begin{algorithm}[H]
    \caption{\small{: at producer $P$, prepare extra location information (Collaborative architecture) \label{alg:producer-collaborative-architecture}}}
    \begin{algorithmic}[1] 
    \STATE PrepareExtraLocationInformation()
       \STATE $RL,R \leftarrow ComputePollReplies()$    
       \IF {$sum(R) > 1 \AND (\exists! b \in R,$ $b= max (R))$}
                 \STATE $collaborativeLocation \leftarrow X|\{( \textsf{location}, \textsf{value},X), (\textsf{numberOfReplies},\textsf{value}, max(R))\}  \in RL  $ 
                 \STATE $lc \leftarrow \{( \textsf{location}, \textsf{value}, collaborativeLocation), (\textsf{location},\textsf{collaborativelyDecided}, true)\}  $
                 
       \ELSE \STATE $lc \leftarrow \{( \textsf{location}, \textsf{value}, locallyProposedLocation), (\textsf{location},\textsf{collaborativelyDecided}, false)\}  $
       \ENDIF
       \RETURN $lc$

    \end{algorithmic}
\end{algorithm}

Notifications with a location not collaboratively verified are not certified by the broker. In this architecture, the role of the broker is reduced to checking the collaborative verification indication and certifying the notifications accordingly (Algorithm~\ref{alg:broker-collaborative-architecture}). It is responsible for providing location certifications but does not decide when to provide them. 

The local broker is assumed to truthfully send the location agreed in the poll. However, when~polled by a neighbor, it answers with the locally proposed location value, which may not be true.

\begin{algorithm}[H]
    \caption{\small{: at broker $B$, verify location (Collaborative architecture) \label{alg:broker-collaborative-architecture}}}
    \begin{algorithmic}[1] 
    \STATE VerifyLocation(Notification $n$)
       \STATE locCertified  $\leftarrow \{Y | (\textsf{location},\textsf{collaborativelyDecided},Y) \in n\}$
       \STATE remove triple $(\textsf{location},\textsf{collaborativelyDecided},*)$ from $n$
       \RETURN locCertified 
    \end{algorithmic}
\end{algorithm}

\section{Simulation}\label{sec:simulation}
\begin{figure}[H]
\centering
\includegraphics[width=0.9\textwidth]{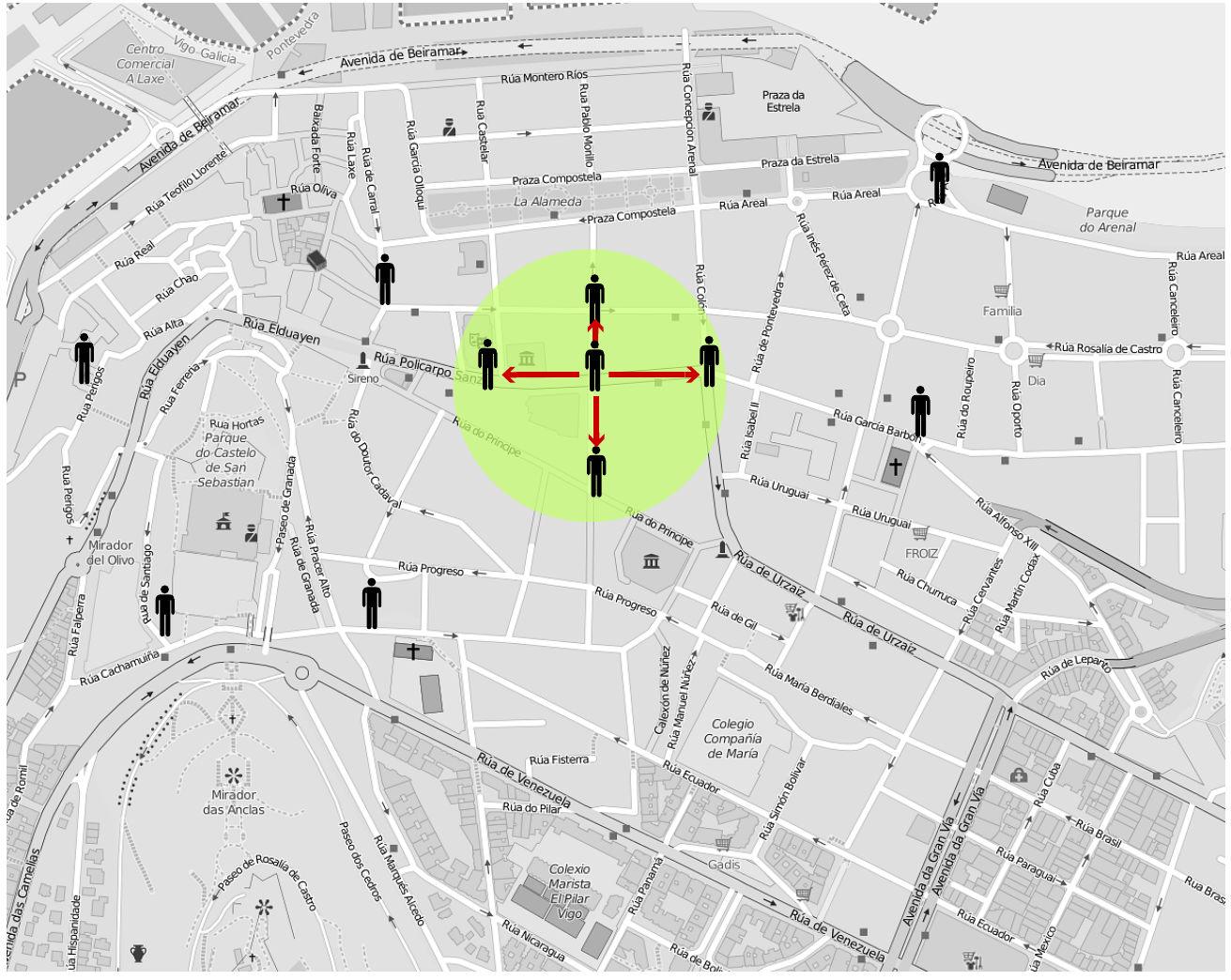}
\caption{Urban area used in simulation. An example of the proximity area around a producer is represented with a green circle.\label{figure:map}}
\end{figure}

To evaluate our architectures, we relied on network simulation. We simulated the noise-measurement scenario described in Section~\ref{sec:scenario}. To do so, we employed realistic pedestrian traces based on a real urban map.
We  chose the city center of Vigo (Spain), an area of 1.46 km$^2$, where most of the streets are either pedestrian-only or include wide sidewalks, which makes it appropriate for pedestrian simulation. Figure \ref{figure:map} depicts the selected area. As an example, the proximity area around a producer is represented by a green circle.

We generated pedestrian traces for 100 pedestrians using the Bonnmotion~\cite{aschenbruck2010bonnmotion} mobility scenario generation tool and a map of the selected area exported from OpenStreetMap~\cite{OpenStreetMap2015}. The traces were generated according to MSLAW~\cite{schwamborn2013introducing}, a  map-based statistical mobility model that extends the Self-similar Least-Action Walk (SLAW) model by including geographic restrictions. As a result, it~targets realistic mobility, considering different mobility metrics ,such as pauses and intercontact time. We generated five different mobility traces. Then, we fed those traces to the ns-3 network simulator~\cite{ns3}, where we implemented our proposed architectures.
We chose ad hoc WiFi as the transmission technology for short-range communication since its transmission range ($\approx$100 m) makes it suitable for urban scenarios. Moreover, we modeled cloud communication through LTE connections. 

To implement our verification strategies, our area was divided into smaller cells. Each of them was assigned a cell identifier, which is the location value producers use as location attribute in their publications. We assumed producers are always aware of what  their position and the identifier of the cell they are in.

Our proposed architectures aim at correctly identifying the veracity of the location claims included in publications at minimum cost. This cost is considered in terms of required infrastructure and consequences in publication
delivery. Thus,  assessment  focused on  appropriate location certifications and at the ability of the system to handle producers' mobility. In that respect, we  carried out simulations where producers choose to provide a fake location according to a certain probability ($P_f$), which follows a uniform distribution.

For the sake of simplicity and without impairing correctness, we did not include consumers in our simulation scenarios. Since we aimed to assess publication creation in producers and publication certification in brokers, broker-consumer communication is out of the scope of this work.

The publication interval was  set to 1 min, which led to frequent sampling times without excessively draining mobile devices' battery. 

Table~\ref{table:simulation-parameters} recaps the main parameters of the simulation that are common to the three architectures. This section continues with the description of the implementation details specific to each architecture. 

\begin{table}[H]
\centering
\caption{Shared simulation parameters.}\label{table:simulation-parameters}
\begin{tabular}{l l }
    
    \hline
   \textbf{ Short-range communication }& WiFi ad hoc mode \\
    & Connection type: Direct \\
    & Connection pattern: Random \\ \hline
    \textbf{Cloud communication} & LTE \\ \hline
    \textbf{Scenario} & Number of producers: 100\\
    & Simulation duration: 1 h\\
    & Simulation area: 1.46 km$^2$\\ \hline
    \textbf{Mobility} & Model: MSLAW\\ 
     & Speed: 0.9--1.5 m/s\\ 
     & Pause time: 10--50 s \\ \hline
    \textbf{Parameters} & Notification interval: 1  min \\ \hline
\end{tabular}

\end{table}
 
\subsection{Fixed-Brokers Architecture}
This simulation scenario included mobile producers and fixed brokers, all of them equipped with ad hoc WiFi. The area was divided in a grid of square of cells of $\approx$200 $\times$ 200~m. There is a fixed broker in the middle of every cell. A number identifier is assigned both the broker and the cell. When they issue a publication, producers include as a location tag the cell identifier that corresponds to their location coordinates. Brokers provide a location certification to every publication they receive whose cell identifier corresponds to their own.

Producers establish a connection with a broker from whom they have received a broker beacon when they are not previously connected to any other. When connected, they send their periodical publications to their broker. If not connected, notifications are queued in the producer.

Our work focused on designing a communication strategy that could allow mobile producers to stay connected most of the time to their nearest broker despite their continuous movement. Thus, the number of publications lost and not transmitted are minimized while the number of publications whose location claim is correctly assessed is maximized. With this goal in mind, extensive simulations are performed testing different time values for broker beacon interval and maximum connection time. The latter is the time after which producers consider themselves disconnected from a certain broker if they have ceased to receive beacons. It was found that the publication delivery rate improves when the maximum connection time matches the broker beacon interval. Moreover,  for pedestrian mobility, this time is best set to 2~s. 

\subsection{Assigned-Brokers Architecture}
This simulation scenario included mobile producers that communicate with brokers in the cloud using LTE technology. Producers act as LTE user equipment (UE) and connect to a base station (eNB) that covers the whole simulation area. Producers are also equipped with ad hoc WiFi, which allows them to interact with other nearby producers. 

There is a broker assigned to every one of the cells in the area, which is in charge of managing the publications issued by producers there. Producers are aware of the cell they are in and include the cell identifier in the publications they issue. Moreover, they also include a list of their neighbor producers, which they have detected in their surroundings by using periodic ad hoc WiFi beaconing. Producers are assumed to always provide their true neighbor list, even when they lie about their location. 

The size of the cells has been decided as a function of ad hoc WiFi transmission range ($\approx$100~m). The size in the first architecture ($\approx$200 $\times$ 200~m) allows fixed brokers to cover the whole area from their position in the middle. However, in this scenario, areas need to be smaller ($\approx$100 $\times$ 100~m) so that producers are able to communicate with most of their peers in the area. As a result, producers are also more likely to receive beacons from their neighbor areas.

If periodic beaconing between producers is more frequent than publications, producers can detect changes in their surroundings faster than the broker. This may result in incorrect publication certification. As a result, we  set the publication interval much lower than in the previous simulation, as low as the beaconing interval. To have a fair comparison with the first architecture, notifications shown in the results section were  sampled every minute. 

The simulation includes a warm-up period where notifications are not accounted for and producers provide only true locations. Thus, false location claims are always dealt with in a steady state. This period was  set to 30~s. Since the beaconing and notification intervals were each set to 2~s, this was enough for sufficient notifications to be processed.

\subsection{Collaborative Architecture}
The simulation setup is similar to the one in the assigned-brokers architecture: producers act as LTE UE connected to an eNB that covers the simulation area. However, the grid distribution adopted is the same as in the fixed brokers architecture, where the area is divided in a grid of 200 $\times$ 200~m cells. This cell size offers an appropriate balance between location precision and the reduced likelihood of receiving too many poll replies from neighbor areas. Producers can be connected to any access broker regardless of their position.

Moreover, producers communicate with their neighbors using ad hoc WiFI when polling or answering a poll. There is no beaconing in this architecture.

In collaborative location assessment, producers are in charge of verifying location claims without brokers being involved. Given that local producers independently decide to provide a false location claim, this strategy makes it unlikely that a notification with a false claim is certified. For this to happen, at least two neighbor producers have to provide the same fake claim. Moreover, the rest of the poll participants (if any) have to provide location claims that either do not coincide with each other or do not add up to more than the fake location claims. 
Waiting time for poll answers was  set to 2~s, enough time for producers to receive their peers' answers without too much delay in the publication~process.

\section{Evaluation}\label{sec:evaluation}
 To assess our architectures, we examined the number of publications, how they are transmitted, and how they are evaluated for certification. Thus, we evaluated the number of publications that are certified and the cost of such certification in every one of the architectures in terms of lost and incorrectly certified and uncertified publications. 
 
For every architecture, we  ran simulations where producers never provided fake location claims ($P_f = 0$), and simulations where they could provide a fake claim ($P_f = 0.3$). The simulation results in Tables~\ref{table:simulation-5} and \ref{table:simulation-6} show the average results obtained after simulating every architecture with five different trace files. 
The tables include a row for every verification strategy proposed in Section~\ref{sec:architecture}: one for the fixed-brokers architecture, two for the assigned-brokers architecture, and one for the collaborative architecture.  The tables are divided into three parts. Subtable (a) is the flow of notifications in the producer and through the network, while Subtables (b) and (c)  depict the notifications that are delivered to an access broker and how they are processed, respectively. These subtables display the treatment of the notifications with true and false location claims, respectively. For consistency, the tables include the same columns for the different verification strategies and probabilities of fake location claims ($P_f$). However, it should be noted that some of the columns are only relevant to a certain strategy or~probability.

\begin{table}[H]
\centering
\caption{Simulation statistics $P_f = 0$.\label{table:simulation-5}} 
\caption*{(a) Producer and network flow \label{subtable:first}}
\begin{tabular}{l c c c c c c c c}
    
     \hline
    \multicolumn{1}{c}{} & \multicolumn{3}{c}{\textbf{Published}} & \multicolumn{2}{c}{\textbf{Sent}}& \multicolumn{1}{c}{\textbf{Queued}}& \multicolumn{1}{c}{\textbf{Lost}} & \multicolumn{1}{c}{\textbf{Delivered}} \\ \hline
   \multicolumn{1}{c}{} & \multicolumn{1}{c}{\textbf{Total}} & \multicolumn{1}{c}{\textbf{True loc}} & \multicolumn{1}{c}{\textbf{False loc}} & \multicolumn{1}{c}{\textbf{True loc}} & \multicolumn{1}{c}{\textbf{False loc}} & \multicolumn{3}{c}{} \\ \hline
	Fixed & 5,900 & 5,900 & 0 & 5,900 & 0 & 0.2 & 62.6 & 5,837.2 \\
Assigned-CLS & 5,900 & 5,900 & 0 & 5,900 & 0 & 0 & 0 & 5,900 \\
Assigned-NLS & 5,900 & 5,900 & 0 & 5,900 & 0 & 0 & 0 & 5,900 \\
Collaborative & 5,900 & 5,900 & 0 & 5,573.4 & 326.6 & 0 & 0 & 5,900 \\
     \hline
\end{tabular}

\caption*{(b) Delivered notifications with a true location claim \label{subtable:second}}
\begin{tabular}{l c c c c c}
    
   \hline
     \multicolumn{1}{c}{}  &\multicolumn{5}{c}{\textbf{Delivered}} \\ \hline
     \multicolumn{1}{c}{} & \multicolumn{5}{c}{\textbf{True location}} \\ \hline
     & \textbf{Total} &  \textbf{Cert} & \textbf{\%} & \textbf{Uncert} & \textbf{\%} \\ \hline

Fixed & 5,837.2 &  5,275.8 & 90.38 & 561.4 & 9.62 \\
Assigned-CLS & 5,900 & 4,005.4 & 67.89 & 1,894.6 & 32.11 \\
Assigned-NLS & 5,900 & 631.8 & 10.71 & 5,268.2 & 89.29 \\
Collaborative & 5,573.4 & 4,073 & 73.08 & 1,500.4 & 26.92\\
     \hline
\end{tabular}

\caption*{(c) Delivered notifications with a false location claim \label{subtable:third}}
\begin{tabular}{l c c c c c}
    
   \hline
     \multicolumn{1}{c}{}  &\multicolumn{5}{c}{\textbf{Delivered}} \\ \hline
     \multicolumn{1}{c}{} & \multicolumn{5}{c}{\textbf{False location}} \\ \hline
     & \textbf{Total} &  \textbf{Cert} & \textbf{\%} & \textbf{Uncert} & \textbf{\%} \\ \hline
Fixed & 0 & 0 & 0 & 0 & 0 \\
Assigned-CLS & 0 & 0 & 0 & 0 & 0 \\   
Assigned-NLS & 0 & 0 & 0 & 0 & 0\\
Collaborative & 326.6 & 326.6 & 100 & 0 & 0 \\  
     \hline
\end{tabular}

\end{table}

Subtable (a) starts with the number of notifications created by the producers (i.e.,~\textit{Published}) and describes the distribution of true and false location claims between those. In~the collaborative architecture, the notifications are processed by the local broker before being sent into the network. As a result of the collaborative verification process, their location claims may change and therefore, the distribution of true and false locations claims can be different. This distribution is reflected in the \textit{Sent} column. Sent notifications may not be delivered to the access broker, they may remain in the producers' queue at the end of the simulation (\textit{Queued}) or they may simply not reach the broker due to network issues (\textit{Lost}). Subtable (b) displays the number of notifications delivered to an access broker that include a true location claim and how many of them are certified and uncertified by the broker. The rates of certified and uncertified delivered notifications with a true location claim are also expressed as a percentage. Subtable (c) is identical to the second,  but considers delivered notifications with a false location claim instead.

\begin{table}[H]
\centering
\caption{Simulation statistics $P_f = 0.3$.\label{table:simulation-6}} 
\caption*{(a) Producer and network flow \label{subtable:first}}
\begin{tabular}{l c c c c c c c c}
    
     \hline
    \multicolumn{1}{c}{} & \multicolumn{3}{c}{\textbf{Published}} & \multicolumn{2}{c}{\textbf{Sent}}& \multicolumn{1}{c}{\textbf{Queued}}& \multicolumn{1}{c}{\textbf{Lost}} & \multicolumn{1}{c}{\textbf{Delivered}} \\ \hline
   \multicolumn{1}{c}{} & \multicolumn{1}{c}{\textbf{Total}} & \multicolumn{1}{c}{\textbf{True loc}} & \multicolumn{1}{c}{\textbf{False loc}} & \multicolumn{1}{c}{\textbf{True loc}} & \multicolumn{1}{c}{\textbf{False loc}} & \multicolumn{3}{c}{} \\ \hline
   
Fixed & 5,900 & 4,106.8 & 1,793.2 & 4,106.8 & 1,793.2 & 0.2 &  62.6 & 5,837.2\\
Assigned-CLS & 5,900 & 4,088.2 & 1,811.8 & 4,088.2 & 1,811.8 & 0 & 0 & 5,900\\
Assigned-NLS & 5,900 & 4,088.2 & 1,811.8 & 4,088.2 & 1,811.8 & 0 & 0 & 5,900\\
Collaborative & 5,900 & 4,137 & 1,763 & 4,534 & 1,366 & 0 & 0 & 5,900 \\

     \hline
\end{tabular}

\caption*{(b) Delivered notifications with a true location claim \label{subtable:second}}
\begin{tabular}{l c c c c c}
    
   \hline
     \multicolumn{1}{c}{}  &\multicolumn{5}{c}{\textbf{Delivered}} \\ \hline
     \multicolumn{1}{c}{} & \multicolumn{5}{c}{\textbf{True location}} \\ \hline
     & \textbf{Total} &  \textbf{Cert} & \textbf{\%} & \textbf{Uncert} & \textbf{\%} \\ \hline
Fixed & 4,062.2 & 3,670 &  90.35 & 392.2 & 9.65 \\
Assigned-CLS & 4,088.2 & 2,542.6 & 62.19 & 1,545.6 & 37.81 \\
Assigned-NLS & 4,088.2 & 520 & 12.72 & 3,568.2 & 87.28 \\
Collaborative & 4,534 & 3,220 & 71.02 & 1,314 & 28.98 \\
     \hline
\end{tabular}

\caption*{(c) Delivered notifications with a false location claim \label{subtable:third}}
\begin{tabular}{l c c c c c}
    
   \hline
     \multicolumn{1}{c}{}  &\multicolumn{5}{c}{\textbf{Delivered}} \\ \hline
     \multicolumn{1}{c}{} & \multicolumn{5}{c}{\textbf{False location}} \\ \hline
     & \textbf{Total} &  \textbf{Cert} & \textbf{\%} & \textbf{Uncert} & \textbf{\%} \\ \hline
     
     Fixed & 1,775 & 0 & 0 & 1,775 & 100 \\
Assigned-CLS & 1,811.8 & 238.4 & 13.16 & 1,573.4 & 86.84 \\
Assigned-NLS & 1,811.8 & 2.2 & 0.12 & 1,809.65 & 99.88 \\
Collaborative & 1,366 & 366 & 26.79 & 1,000 & 73.21 \\    
     
     \hline
\end{tabular}

\end{table}

The number of published notifications is the same for all the architectures while, due to their random generation, the number of publications with a false location claim slightly differs. The number of sent publications  only differs with the number of published ones in the collaborative architecture, where the location is the result of the neighbor poll and not directly the value decided by the producer itself. Publications are only queued or lost in the fixed-brokers architecture, as explained in Section~\ref{subsubsec:fixed-issues}. In the other two architectures, producers communicate with their brokers using LTE technology and, since LTE coverage of the area is total, producers are always able to send their publications. Therefore, there are neither queued nor lost notifications.

All the architectures presented solid results in detecting false location claims. However, their~simulation outcomes differed. In this section, we examine the statistics of each architecture simulation in detail.

\subsection{Fixed-Brokers Architecture}
In this architecture, the average rate of lost messages  was slightly over 1\% (Tables~\ref{table:simulation-5}a and \ref{table:simulation-6}a, 62.6/5900 $=$ 1.061\%) while there were practically no queued messages. Thus, in the worst case, at least 98\% of all the publications were received at an access broker and transmitted. Results were also good for the correct certification rate. Every message with a false location was correctly identified and therefore none of them was certified (Tables~\ref{table:simulation-5}c and \ref{table:simulation-6}c) . In other words, 100\% of messages with a false location were successfully assessed. The rate of messages incorrectly uncertified did not reach 10\% (Tables~\ref{table:simulation-5}b and \ref{table:simulation-6}b). As a consequence, more than 90\% of messages with true location were correctly identified. In~general terms, we can conclude that the system reasonably meets its target of delivering publications to the broker and correctly providing location certifications.

\subsubsection{Publication and Certification Issues}\label{subsubsec:fixed-issues}
Due to producer mobility, it is difficult to completely solve publication loss and certification issues.  There exist queued publications that never leave their source producer during  simulation time and are, therefore, not delivered to any broker. These publications may have been transmitted if the simulations were longer. Lost publications are those that have been sent by a producer but they have not reached an access broker. In this architecture, the certification issue affects mostly publications with a true location claim that are not certified due to their mobility.

\begin{figure}[H]
\centering
\begin{minipage}[b]{0.3\textwidth}
\centering
\includegraphics[width=0.9\textwidth]{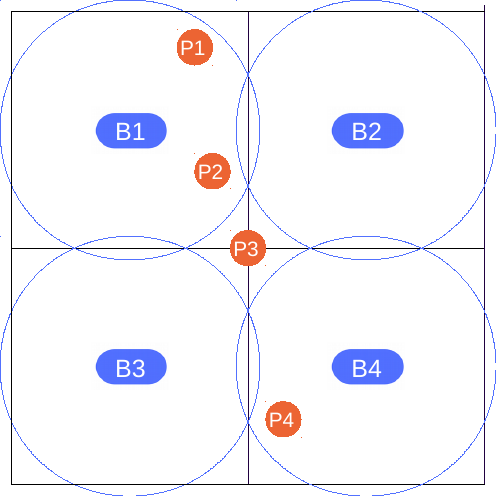}
\caption{Producers at time \textit{t}\label{subfig:nodes-before}}
\end{minipage}%
\begin{minipage}[b]{0.3\textwidth}
\centering
\includegraphics[width=0.9\textwidth]{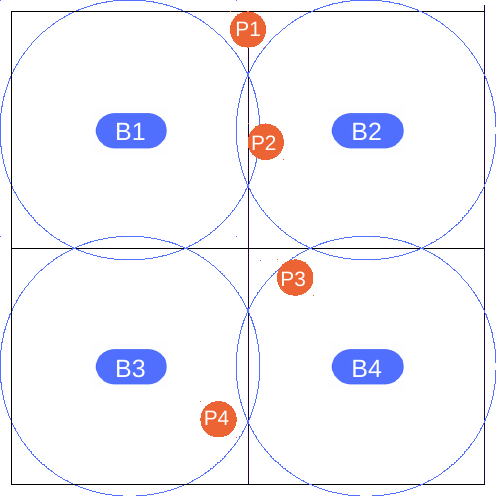}
\caption{Producers at time \textit{t + 1s}\label{subfig:nodes-after-1}}
\end{minipage}%
\begin{minipage}[b]{0.3\textwidth}
\centering
\includegraphics[width=0.9\textwidth]{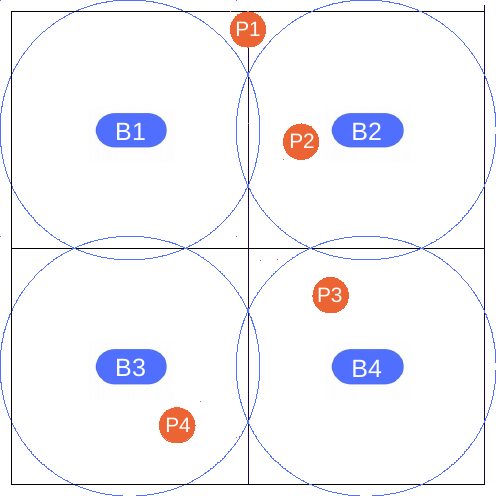}
\caption{Producers at time \textit{t + 2s}\label{subfig:nodes-after-2}}
\end{minipage}%
\vspace{-6pt}

\caption{Example of area grid and producers' movement.\label{fig:example-nodes}}
\end{figure}

Figure~\ref{fig:example-nodes} illustrates situations where publications are lost or incorrectly uncertified. It represents a grid divided into four different areas, with a broker placed at the center of each one. The areas were  designed to maximize area coverage with the minimum number of brokers while avoiding overlapping areas. The circles around the brokers represent the area covered by their signal. Each of the subfigures represents a different temporal instant. Thus, all of Figure~\ref{fig:example-nodes} depicts a temporal sequence where producers are moving through the area. We assumed that brokers send beacons at \textit{t} and \textit{t + 2}, and that producers issue a periodic publication at \textit{t + 1}. 

At \textit{t}, Producers 1 and 2, being in the same area as Broker~1,  receive a beacon from this broker and establish a connection with it. Producer~4 connects, in turn, to Broker~4. Producer~3, however, is in an area without coverage so it does not receive any beacon and considers itself as  disconnected. 

At \textit{t + 1}, all the producers have moved. However, they have not received any new broker beacon, and neither have they missed it since the maximum connection time (2~s) has not expired yet. As a result, they will consider their previous situation when delivering a publication. Although Producer~1 considers that it is still connected to Broker~1, it is in an area without coverage and, as a consequence, its publication is lost. The same happens to the event notification delivered by Producer~4 since it is issued to Broker~4, which is not anymore in the transmission range. Producer~2 has also changed area but, since it is on an overlapping edge of the new area, it is still able to communicate with Broker~1. However, its publication includes a location claim of area 2 and, therefore, it cannot be certified. Since the notification includes the actual area where the producer is at the publication time but it is processed by a broker in charge of a different area, Producer~2's publication becomes an uncertified publication with true location. Producer~3 has entered Broker~4's area but, since it has not established any connection, it saves the publication in its internal queue. 

At \textit{t + 2}, most of the producers have moved again except for Producer~1. In doing so, Producer~1 has not received any beacon and has realized that its connection with Broker~1 is broken. Producers 2, 3, and 4 have received beacons from their nearest brokers (Brokers 2, 4, and 3, respectively) and established a connection with them.

The above examples  describe what may happen when nodes are on an area edge. If they leave the area, it takes some time for them to realize that they cannot reach  the broker to which they were connected anymore. Consequently, they may send a publication that never reaches the broker and is then lost (e.g., Producer~1 at \textit{t + 1}). Publications may also become lost due to collisions on the WiFi channel. It is also possible that nodes move to an overlapping area that corresponds to a new broker. In this case, they  send publications to their old broker that include their new location (e.g., Producer~2 at \textit{t + 1}). This is one of the situations where publications with true location are not certified. The other possible scenario takes place when nodes that have lost their connection for some time find another broker. Then, they issue their queued publications to the new broker.

\subsection{Assigned-Brokers Architecture}
Since the verification strategy relies only on mobile producers, simulation results cannot meet the ones obtained in the fixed-brokers architecture. Neighbor changes are quick, even more with smaller areas, and, therefore, it is difficult for  producers to  always be aware of who is nearby.  The~percentage of notifications with a false location claim that are certified is low for both of the verification strategies (Tables~\ref{table:simulation-5}c and \ref{table:simulation-6}c) but is zero in NLS because false location notifications do not include any neighbor that is registered in the broker. Publications that are incorrectly certified in CLS satisfy the conditions required as they did not include any neighbor unknown to the broker nor fail to include a neighbor known to the broker. They are publications that include an empty neighbor list sent to a broker that do not have any producer registered in the area, that is to say, they are labeled with a location claim that belongs to an empty area.

Even though the system correctly identifies most of the notifications with a false location claim, it~does not work that well at recognizing true claims (Tables~\ref{table:simulation-5}b and \ref{table:simulation-6}b). Performance is especially poor in NLS. Node density and the number of cells in the grid play a role in this situation. Due to  cell distribution, there are many more areas than producers. As a result, it is not unlikely that a producer is alone in this cell. Although there is a risk in certifying lonely producers as explained in the previous paragraph, it pays off in terms of general results in a low-density scenario like this.

Nodes providing false claims also interfere in the verification of the publications issued by their neighbors. A publication including a neighbor that has not been registered in the broker results in uncertification in CLS. This may happen when a producer includes an actual neighbor whose publication provides a false claim. When a publication is wrongly certified, the producer is also included in the list of producers registered with the broker assigned to that area. For a certain time, this  prevents true notifications from being certified since the broker does not certify publications that do not include this producer, which is not known by any of the actual producers in the area.

Verification strategies have different strengths and, therefore, can be applied to different scenarios. CLS is better suited for scenarios where the overall performance of the system is considered. In~situations where it is of paramount importance to avoid false location claims, NLS should be employed.  It is successful as long as the assumption of producers sending their true neighbor list is~maintained.

\subsection{Collaborative Architecture}
Since location values in publications may change after  collaborative assessment, the analysis of this architecture is more complex. Tables~\ref{table:simulation-5} and \ref{table:simulation-6}  first show a distribution of true and false location claims before (\textit{Published}) and after (\textit{Sent}) collaborative verification. Then, they present the same distribution once  publications are delivered to the broker (\textit{Delivered}). In the first two architectures, location values in notifications do not change during the verification process. Therefore, this classification is sufficient to understand the operation of the verification strategies. However, in the collaborative case, it is necessary to study how location claims are modified to assess the outcome of the verification strategy.

With that aim, Tables~\ref{table:simulation-8} and \ref{table:simulation-9} provide a more detailed study of the results in Tables~\ref{table:simulation-5} and \ref{table:simulation-6}. First, Table~\ref{table:simulation-8} presents the number of delivered notifications and the rates of certification. Then, Table~\ref{table:simulation-9} takes a closer look at the veracity of the location claims of certified and uncertified notifications and classifies them according to whether they have been modified in the verification process.


\begin{table}[H]
\centering
\caption{Certification statistics for delivered notifications in the simulation of the collaborative architecture.}\label{table:simulation-8}
\begin{tabular}{c c c c c c}
    
    \hline
\multicolumn{6}{c}{\textbf{Delivered} } \\ \hline \multicolumn{1}{c}{} & \multicolumn{1}{c}{\textbf{Total}} & \multicolumn{2}{c}{\textbf{Certified}} & \multicolumn{2}{c}{\textbf{Uncertified}} \\ \hline    
$\mathbf{P_f}$  & & \textbf{Total} & \textbf{\%} & \textbf{Total} & \textbf{\%} \\ \hline
0 & 5,900 & 4,399.6 &  74.57 & 1,500.4 & 25.43 \\ 
0.3 & 5,900 & 3,586 & 60.78 & 2,314 & 39.22 \\
    \hline
\end{tabular}

\end{table}

Location claims in certified notifications have always been collaboratively decided. As a result, the original claim inserted by the producer may have been replaced with a different value that~may have altered the veracity of the claim. Consequently, a true location claim can either maintain its value (\textit{Remain true}) or have had it changed for a false one (\textit{True to false}). Likewise, a false location claim may still be false (\textit{Remain false}) or may have been corrected to a valid value (\textit{False to true}). \textit{Remain false} includes both unchanged claims and claims that have been substituted by a different value, also~incorrect. On the other hand, uncertified notifications contain always the original location claims proposed by the producers, i.e., they maintain their initial veracity.

The results show how most of the notifications published are certified (Table~\ref{table:simulation-8}), which means the location they include has been collaboratively assessed. Out of these, most of the notifications included originally a true location that was maintained after the collaborative evaluation or initially presented a false claim that was corrected (Table~\ref{table:simulation-9}a). As a matter of fact, a third of the total publications with false claims were corrected to their true location (598 corrected, 1165 remained false ($165 + 1000$)) while only about 5\% of the true location claims were modified (201 vs. 3936 unchanged ($2622 + 1314$)).


\begin{table}[H]
\centering
\caption{Detailed statistics for delivered notifications in the simulation of the collaborative architecture.\label{table:simulation-9}}

\caption*{(a) Certified notifications}
\begin{tabular}{c c c c c c c c c c}
    
    \hline
   \multicolumn{1}{c}{} & \multicolumn{9}{c}{\textbf{Certified}} \\ \hline
   \multicolumn{1}{c}{\small{$\mathbf{P_f}$}} & \multicolumn{1}{c}{\small{\textbf{Total}}} & \multicolumn{2}{c}{\small{\textbf{Remain true}}} & \multicolumn{2}{c}{\small{\textbf{False to true}}} & \multicolumn{2}{c}{\small{\textbf{Remain false}}} & \multicolumn{2}{c}{\small{\textbf{True to false}}} \\ \hline
 & & \textbf{Total} & \textbf{\%} & \textbf{Total} & \textbf{\%} & \textbf{Total} & \textbf{\%} & \textbf{Total} & \textbf{\%} \\ \hline  
0 & 4,399.6 & 4,073 & 92.58 & 0 & 0 & 0 & 0 & 326.6 & 7.42 \\
0.3 & 3,586 & 2,622 & 73.12 & 598 & 16.68 & 165 & 4.6 & 201 & 5.61 \\  
    \hline
\end{tabular}

\caption*{(b) Uncertified notifications}
\begin{tabular}{c c c c c c}
    
    \hline
     \multicolumn{1}{c}{} & \multicolumn{5}{c}{\textbf{Uncertified}} \\ \hline \multicolumn{1}{c}{\small{$\mathbf{P_f}$}} & \multicolumn{1}{c}{\small{\textbf{Total}}} & \multicolumn{2}{c}{\small{\textbf{Remain true}}}  & \multicolumn{2}{c}{\small{\textbf{Remain false}}} \\ \hline
    & & \textbf{Total} & \textbf{\%} & \textbf{Total} & \textbf{\%} \\ \hline 
0 & 1,500.4 & 1,500.4 & 100 &  0 & 0 \\
0.3 & 2,314 & 1,314 & 56.78 & 1,000 & 43.22 \\
    \hline
\end{tabular}

\end{table}

Every modified claim is certified since all of them are decided through collaborative assessment. Even though there are notifications that remain or become false after the assessment (Table~\ref{table:simulation-9}a), it~should be noted that their locations always correspond to neighbor areas. This way, originally false location claims, which are certified and  not corrected to their true location, are at least changed to an adjacent location claim. This also happens with originally true location claims that are modified to a false value when their producers are on an area limit and, therefore, receive multiple poll replies from a neighbor area. As a result, every false certified notification comes from a producer standing on the edge of an area and consequently close to the neighbor location included in their notification. Consequently, their claims are not useless but give a valid clue about the actual producer location.

Uncertified notifications are those whose collaborative assessment is declared undecided and then include the original location proposed by the producer (Table~\ref{table:simulation-9}b). Two factors contribute to the number of uncertified notifications. The first one is the number of tied polls. When producers receive poll replies from different areas and there is a draw between the results, the notification is uncertified. The percentage of true uncertified notifications increases in the simulation with false notification claims since lying producers may create draws in polls where all the producers are in the same area. Polls with only two participants (i.e., have received only one reply) are especially sensitive to this. In this case, if the two location claims proposed coincide, the poll result is certified, whereas a poll with two different location claims results in uncertification. There is a significant number of minimal polls in the simulation. This is caused by the second factor involved in uncertification: low density of devices. Although 100 nodes is realistic density for a quiet urban area, it may be slightly low for a collaborative scheme like this. In conclusion, notifications from producers that do not receive replies are always uncertified, and notifications from producers that receive only one reply are sensitive to be also uncertified.

\subsection{Discussion}
Although the three architectures can be successfully employed for location verification in DEBS, each of them is more suitable for a different scenario. In order to paint a clear picture of the simulation setup and enable a fair result comparison, the main differences between the architectures are outlined in Table~\ref{table:comparing-architectures}. The table starts by presenting the different cell sizes employed in the architectures. Then, it~indicates which of them require dedicated infrastructure and periodic beacons. Moreover, the table details the different levels of broker involvement and neighbor dependency the architectures entail.

 The fixed-brokers architecture is always the most reliable option and provides the best results despite losses. As a result, it is advisable in scenarios where dependability is crucial and a high investment in infrastructure is possible. Moreover, this architecture does not rely on the existence of neighbor producers and works well with low producer density.

 At the opposite end of the spectrum, the collaborative architecture fits best in flexible scenarios where no resources are required to be allocated for broker assignment. Although polling is costly, it~is not as much as periodic beacons in the other architectures. However, this architecture certifies the most publications with false location claims (Tables~\ref{table:simulation-5}c and \ref{table:simulation-6}c). It should be noted that the location of this publications has indeed been corrected but to an area adjacent to the correct one. As a result, ~information is more useful than the original false location claim. Nevertheless, this architecture is less appropriate in scenarios where precision is essential.
\begin{table}[H]
\centering
\caption{Differences between architectures.}\label{table:comparing-architectures}
\scalebox{.93}[0.93]{
\begin{tabular}{l >{\centering}m{27mm}  >{\centering}m{20mm}  >{\centering}m{20mm}   >{\centering}m{20mm} }
\hline
\multicolumn{1}{c}{} & \multicolumn{1}{c}{\textbf{Fixed Brokers}} &  \multicolumn{2}{c}{\textbf{Assigned Brokers}} & \multicolumn{1}{c}{\textbf{Collaborative}}\\ \hline
\multicolumn{1}{c}{} &  \multicolumn{1}{c}{} & \multicolumn{1}{c}{\textbf{CLS}} & \multicolumn{1}{c}{\textbf{NLS}}& \multicolumn{1}{c}{}\\
\hline

Cell size   & 200 $\times$ 200 m & \multicolumn{2}{c}{100 $\times$ 100 m} & 200 $\times$ 200 m \tabularnewline

Dedicated infrastructure   & Yes & \multicolumn{2}{c}{No} & No \tabularnewline

Intermittent broker connection   & Yes & \multicolumn{2}{c}{No} & No \tabularnewline 

Periodic beaconing  & Yes & \multicolumn{2}{c}{Yes} & No \tabularnewline 

Broker role & Principal & \multicolumn{2}{c}{Intermediate} & Minimal \tabularnewline

Sensibility to producers' density  & None & \multicolumn{2}{c}{High} & High \tabularnewline 

Infrastructure  & Physical & \multicolumn{2}{c}{Cloud} & None \tabularnewline 

Reliability  & High & High & Intermediate & Intermediate \tabularnewline 
\hline
\end{tabular}}
\end{table}

The first verification strategy of the assigned-brokers architecture (CLS) achieves moderately better results by introducing the burden of  broker allocation. NLS is the best alternative when infrastructure deployment is not possible and the system is less tolerant to incorrectly certified claims. If the system has a tolerance, CLS and the collaborative architecture present a better percentage of correctly certified notifications since they work better with true location claims. 

Due to the different infrastructure requirements of the architectures, it is not possible to run different use cases in parallel. The possibility of changing between the architectures is not supported either. The only flexible solution for scenarios with changing circumstances is the implementation of the assigned brokers architectures, which allows switching between the two verification strategies. In any case, the deployment of the system should be preceded by a careful study to determine the requirements of the scenario in order to select the most suitable architecture.

The architectures were evaluated according to the threat model in Section~\ref{subsec:threat-model}, in which there exist location-spoofing attacks where producers act independently. None of the architectures can resist a more complex attack scenario where the brokers are also compromised. However, their behavior in case of colluding producers varies. The assigned-brokers architecture and the collaborative architecture have high neighbor dependency and are especially sensitive to this kind of attack. In contrast, the~fixed-brokers architecture would not be affected at all.

\section{Related Work}
Due to the rise of participatory sensing schemes, it has become necessary to develop solutions to verify the quality and correctness of the contributed data. Data-verification strategies can be general to cover any kind of information, or more specific, targeted at a certain type, such as location.

An example of general data verification is SHIELD~\cite{gisdakis2015shield}, a centralized system based on evidence handling and data mining to identify and filter out erroneous user contributions. However, the~requirement for a bootstrapping phase makes data mining unsuitable for a dynamic distributed scenario like ours.

Moving away from classification and training, Luo and Zeynalvand~\cite{luo2017reshaping} propose a cross-validation framework that recruits crowdworkers to act as data verifiers. Their role is not to duplicate a sensing task, but to assess the credibility of the gathered data. This approach is also centralized and, unlike ours, relies on verifiers with no direct knowledge about the information. 

The reason why location verification has attracted significant research attention is twofold. On~the one hand, location information plays an important role in contextualizing contributed information. On the other hand, location-based services require users to provide a valid location value in order to gain access. The proposals reviewed next are targeted at these aims.

Lederer et al.~\cite{lederer2008collaborative} introduce a certification approach at network level where intermediate nodes tag data through the network with belief ratings based on observed past traffic patterns and source location claims. Relying on previous knowledge information about routing paths is a valid strategy for distributed systems where nodes do not move often and, therefore, paths are not likely to change. As a result, this strategy is not appropriate for a highly dynamic scenario like the one we have considered.

Most location verification mechanisms rely on the collaboration of neighbor devices. Although not specifically labeled as fog architectures, these strategies consist of cloud systems that leverage neighbor information. Saracino et al.~\cite{saracino2018practical} propose a reputation-based system where some of the devices in each area act as hotspots to communicate with their peers in WiFi range. A centralized platform is in charge of receiving the lists of devices that can be reached and update reputation calculations accordingly. 
ILR~\cite{talasila2013improving} aims at detecting false location claims in a centralized crowd-sensing system where data providers are assigned sensing tasks. When submitting the required data, they also include a Bluetooth scan of their surroundings. False location claims are identified once the tasks are received, by comparing Bluetooth scans and also through image processing when the task includes picture~taking.

Unlike ILR, other mechanisms are not postprocessing but designed to be used at runtime. LINK~\cite {talasila2010link} is targeted at location verification for third-party location-based services. It relies on a Location Certification Authority (LCA) that receives information on location claimers and their Bluetooth neighbors. Then, it issues a claim decision to the service for which a claimer wants to authenticate.

APPLAUS~\cite{zhu2013toward}, STAMP~\cite{wang2016stamp}, PROPS~\cite{gambs2014props}, and SPARSE~\cite{nosouhisparse} share LINK's aim but include extra features to deal with privacy and colluding neighbors. Instead of exchanging real identifiers, devices in APPLAUS~\cite{zhu2013toward} use pseudonyms and change them periodically. PROPS~\cite{gambs2014props} uses encryption to preserve the identity of the claimer and its neighbors. Regarding collusion prevention, STAMP~\cite{wang2016stamp} pays attention to transaction history between two users to detect if they appear to have always the same neighbors. In SPARSE~\cite{nosouhisparse}, the central entity of the system decides which of the neighbors should act as location witnesses, avoiding that location claimers can select them themselves.

All these location verification schemes that involve neighbor collaboration are centralized and aimed at sporadic location verification. Therefore, they have not been designed to handle continuous sensing in a very distributed setting.

\section{Conclusions and Future Work} \label{sec:conclusion}
In this paper, we demonstrated the potential of fog-based architectures in the implementation of participatory urban DEBS with location verification support. We  proposed three different architectures: the fixed-brokers architecture, the assigned-brokers architecture, and the collaborative architecture. We have proven them to successfully deal with location verification in realistic urban scenarios, and we have discussed which situations would best benefit from each one of them.

Our simulation results show that our architectures succeed at identifying
many of the false location claims (from 73\% to 100\%) and that, therefore, they support location-spoofing attacks in which producers act independently. Nonetheless, there is room for improvement. It is necessary to refine the behavior of the system on area edges and to improve the mobility between areas in the fixed-brokers architecture in order to reduce publication and certification issues. Additionally, it would be valuable to enhance the assigned-brokers and the collaborative architecture to strengthen their performance in low-density scenarios.

Due to the increasing adoption of mobile sensor-enabled devices, the number of services oriented at them or exploiting their possibilities will continue to grow over the next few years. As a result, the~demand for location verification mechanisms targeted at them will also escalate. Our work is aimed at verifying producer location, but it can easily be adapted for consumers. Thus, event subscriptions could be restricted to consumers in a particular area. This first perspective of our work is related to our previous work in multiscale DEBS~\cite{conan2017multiscale}.

For simplicity, we  worked on the assumption that using short-range communication technology is enough to ensure proximity. However, this may not be true. In order to enhance our system and prevent relay attacks, our architectures could be reinforced with the implementation of a distance bounding protocol~\cite{brands1993distance}. By computing the delay on the exchange of several messages, these~cryptographic protocols establish an upper bound distance for a certain producer. If this distance is larger than a certain limit, peer producers and fixed brokers do not consider this producer as a neighbor. Due to the message exchange and subsequent processing, which includes the application of cryptographic functions, this security improvement comes at a significant computational cost. 

Proximity-based collaboration between peers can be extended to verify additional context-related information. A direct extension of our solution lies in the adaptation of the collaborative architecture. Nodes can share their sensor readings and collaboratively decide a value to be provided to the system. In the other architectures, the extension requires  brokers to implement specific mechanisms to process the publications and decide on their veracity. Thus, the complexity of event processing in the broker is increased. Moreover, other context-related information different from temporal and spatial details are mostly application-dependent. Therefore, these collaborative solutions, albeit useful, would~be less general and restricted to specific scenarios. As a second perspective, in order to extend the functionality of the system, this work could benefit from analyzing group mobility~\cite{sudhakar2017spatial}.

\bibliographystyle{elsarticle-harv}

\end{document}